\documentclass{IEEEtran}
\usepackage{cite}
\usepackage{amsmath,amssymb,amsfonts}
\usepackage{graphicx}
\usepackage{textcomp,nicefrac}
\usepackage{subfigure}
\def\BibTeX{{\rm B\kern-.05em{\sc i\kern-.025em b}\kern-.08em
T\kern-.1667em\lower.7ex\hbox{E}\kern-.125emX}}
\begin{document}
\title{PulseDL-II: A System-on-Chip Neural Network Accelerator for Timing and Energy Extraction of Nuclear Detector Signals}
\author{Pengcheng Ai, Zhi Deng, \IEEEmembership{Senior Member, IEEE}, Yi Wang, Hui Gong, Xinchi Ran and Zijian Lang
\thanks{Manuscript received August 18, 2022; revised November 7, 2022.}
\thanks{This work was supported in part by the National Key Research and Development Program of China (under grant no. 2020YFC01220002), in part by the National Key Research and Development Program of China (under project no. 2020YFE0202001), and in part by China Postdoctoral Science Foundation (under grant no. 2021M690088).}
\thanks{The authors are with the Key Laboratory of Particle and Radiation Imaging (MOE), Department of Engineering Physics, Tsinghua University, Beijing, 100084, China (e-mail: dengz@mail.tsinghua.edu.cn).}
\thanks{Pengcheng Ai is now with the PLAC, Key Laboratory of Quark \& Lepton Physics (MOE), Central China Normal University, Wuhan, 430079, China (e-mail: pengcheng.ai@mails.ccnu.edu.cn).}
}

\maketitle

\thispagestyle{empty} 
\pagestyle{empty}     

\begin{abstract}
Front-end electronics equipped with high-speed digitizers are being used and proposed for future nuclear detectors. Recent literature reveals that deep learning models, especially one-dimensional convolutional neural networks, are promising when dealing with digital signals from nuclear detectors. Simulations and experiments demonstrate the satisfactory accuracy and additional benefits of neural networks in this area. However, specific hardware accelerating such models for online operations still needs to be studied. In this work, we introduce \emph{PulseDL-II}, a system-on-chip (SoC) specially designed for applications of event feature (time, energy, etc.) extraction from pulses with deep learning. Based on the previous version, \emph{PulseDL-II} incorporates a RISC CPU into the system structure for better functional flexibility and integrity. The neural network accelerator in the SoC adopts a three-level (arithmetic unit, processing element, neural network) hierarchical architecture and facilitates parameter optimization of the digital design. Furthermore, we devise a quantization scheme compatible with deep learning frameworks (e.g., TensorFlow) within a selected subset of layer types. We validate the correct operations of \emph{PulseDL-II} on field programmable gate arrays (FPGA) alone and with an experimental setup comprising a direct digital synthesis (DDS) and analog-to-digital converters (ADC). The proposed system achieved 60 ps time resolution and 0.40\% energy resolution at signal to noise ratio (SNR) of 47.4 dB.
\end{abstract}


\section{Introduction}
\label{sec:introduction}
\IEEEPARstart{A}{dvances} in electronic design and manufacturing technology have greatly influenced the methodology and sensitivity of nuclear detectors. High-speed digitizers~\cite{AMELI2019286} with associated processing \& storage circuitry revolutionized the front-end electronics (FEE), especially in performance-critical conditions. For system designers, however, the evolution is a double-edged sword: for one thing, it provides an abundance of raw data for experimenting various feature extraction algorithms (both traditional and intelligent); for another, it poses considerable pressure on the readout system, especially hardware and software capacities within the data link. It is an important issue to maximize the benefits of signal sampling systems while keeping the data throughput controllable and the overall complexity acceptable.

Deep learning~\cite{LeCun2015}, essentially multi-layer neural networks with delicate structures, developed dramatically in 2010s as a frontier in machine learning and artificial intelligence. Big data from nuclear detectors made it possible to apply this novel method to many tasks directly related to detector signals. For example, in high energy physics (HEP), deep neural networks have been applied for particle identification~\cite{9448194}, particle discrimination~\cite{9459756} and also low-level triggers~\cite{9445104}. In nuclear medicine, convolutional neural networks have been used for accurate estimation of time-of-flight in positron emission tomography detectors~\cite{Berg_2018,Kwon2021}. In plasma physics, deep variational autoencoders have been utilized to integrate diagnostic data of soft X-ray images~\cite{9481901}. Recently, additional merits and understandings about neural networks in nuclear electronics have been explored, such as estimation of heterogeneous uncertainty of nuclear detector signals~\cite{Ai_2022}, and computation of the Cram\'er Rao lower bound of timing to find out limits for neural networks~\cite{Ai_2021}.

The needs of pushing neural networks to front-end arise from the trade-off between data transmission and data processing. For HEP experiments, there are some low-level tasks with online processing demand and stringent latency requirements. In~\cite{9447722}, an application-specific integrated circuit (ASIC) is developed for accelerating the inference of a compact autoencoder model to ease data compression for the high-granularity endcap calorimeter. In~\cite{9445104}, a dense neural network on an FPGA is designed for level 1 trigger directly generated upon raw data. Since the data interfaces and targets in these applications are well defined, a brute-force approach is adopted to implement the network model. In this approach, only weights in the neural network can be adjusted; the network architecture is unchangeable once the design is consolidated and optimized.

Beyond applications above, there are also some scenarios where we intend to fulfill the advantage of neural networks while keeping more flexibility. Neural network accelerators based on the array of processing elements (PE) serve this purpose~\cite{7738524,8686088}. However, the operation of PE-based accelerators relies on the transactions between accelerators and a host processor. This is easily available in sophisticated computer systems, but not in HEP front-end electronics. SoC digital design with microcontrollers/microprocessors integrated can efficiently schedule transactions related to the neural network accelerator and thus make the system autonomous and independent. The SoC neural network accelerator is among a future upgrade of FEE for the electromagnetic calorimeter in the NICA-MPD experiment (Section \ref{sec:elec-cal-at-nica-mpd}).

\begin{figure}[t]
\centerline{\includegraphics[width=0.3\textwidth]{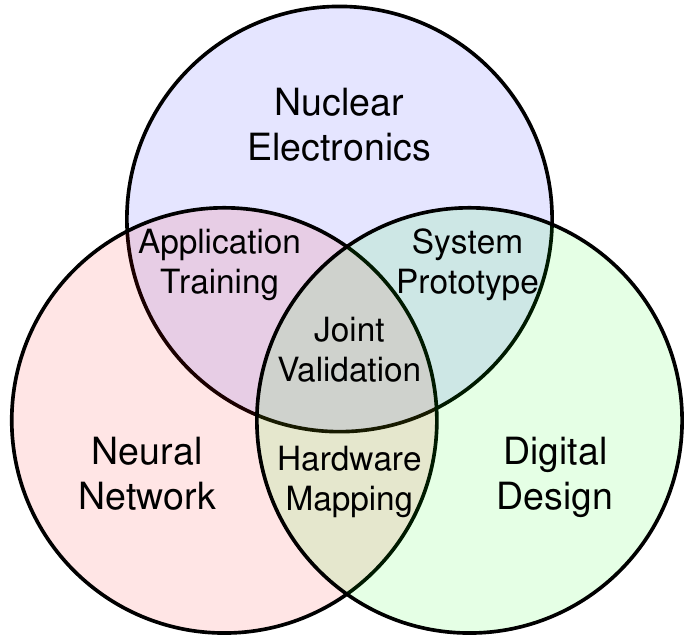}}
\caption{Three elements of design perspectives within the research.}
\label{fig:elements}
\end{figure}

In the following parts of this paper, we will recursively discuss three elements of design perspectives, as shown in \figurename~\ref{fig:elements}. They are:

\begin{itemize}
    \item[$\alpha$] Nuclear Electronics: readout system and signal features.
    \item[$\beta$] Neural Network: architectural research, network training.
    \item[$\gamma$] Digital Design: neural network accelerator and SoC.
\end{itemize}

The three independent elements will produce several intersections, i.e., application training, hardware mapping, system prototype and joint validation. We will use the same notations ($\alpha$, $\beta$, $\gamma$) to relate each section to associated topics.

\section{Electromagnetic Calorimeter at NICA-MPD ($\alpha\gamma$)}
\label{sec:elec-cal-at-nica-mpd}

The Multi-Purpose Detector (MPD) at the NICA collider is designed for protons/heavy ions collisions to study the basic quantum chromodynamics structure of matter~\cite{NICA-MPD-CDR}. The electromagnetic calorimeter (ECAL) in this detector and its FEE are described in~\cite{NICA-MPD-ECAL-TDR}. More recent developments and beam tests of ECAL can be found in~\cite{Shen_2019,LI2020162833}. Silicon photomultipliers coupled to the shashlik-type calorimeter were used to transform scintillation light into current pulses, which would later be transmitted through flat cables. A 64-channel 12-bit commercial ADC board with 62.5 MHz sampling rate digitized the analog signals and sent waveform samples to back-end through optical fibers. The current FEE reached $\sim$1~ns timing resolution and 6\% energy resolution in experimental measurements.

Reaching sub-ns time resolution (i.e., on the order of 150~ps) by ECAL is of great significance for NICA-MPD. With this time resolution, ECAL can work in the time-of-flight (ToF) mode for auxiliary time measurements as an important supplement to the main ToF detector. However, the current FEE is insufficient to support such precision for timing. Real-time transmission of raw ADC samples brings about high data bandwidth and power consumption (about 250 mW/channel). This in return limits the permissible sampling rate of ADC and prevents the FEE from fully realizing the potential resolution.

In our previous work~\cite{Ai_2021}, we have demonstrated that neural networks can effectively achieve near-optimal resolution for feature extraction in the nuclear detector dataflow. Deploying neural networks at FEE is the key point to ensure performance while substantially reducing bandwidth and power consumption. Although commercial neural network accelerators might be a solution to this problem, they do not actually fit into the particular needs of the application scenario. More importantly, they are not efficient in accelerating one-dimensional convolutional neural networks (Section \ref{sec:sig-feature-ext-with-nn}) which are the target workload for nuclear pulse signals.

Therefore, in future upgrade plans of ECAL, digital logic accelerating neural network inferences is proposed to be integrated into FEE for extraction of signal features (time \& energy), along with high-speed pre-amplifiers and ADCs ($\sim$200~MHz). To the best of our knowledge, this is the first implementation of the SoC neural network accelerator for low-level edge-intelligence tasks in high energy physics experiments. The work reported in this paper is aimed at, but not restricted to, application of the neural network accelerator for ECAL at NICA-MPD.

\section{Signal Feature Extraction with Neural Network ($\alpha\beta$)}
\label{sec:sig-feature-ext-with-nn}

\subsection{Building Blocks of Network Structure}

Deep learning models are evolving at a tremendous pace. New elements and architectures are frequently coming out. Here we only focus on one-dimensional (1d) convolutional neural networks~\cite{Ai_2019,Ai_2021}, because they not only fit into the dimensionality of the problem, but also succeed in many machine learning tasks and facilitate parallel computing.

We select four representative building blocks: 1d convolution layer, 1d deconvolution (or transpose convolution) layer, fully-connected layer and nonlinear activation (such as ReLU~\cite{DBLP:conf/iclr/AroraBMM18}). Ref.~\cite{Ai_2021} gives a nice visualization of the former three layers. Regarding the nonlinear activation, it is the key for inductive learning (and thus intelligent signal processing)~\cite{Ye2022}. With nonlinearity, weights in the mapping function are selectively turned off/scaled by the nonlinear function. This subset of neural network layers is supported by mainstream deep learning frameworks, e.g., TensorFlow~\cite{DBLP:journals/corr/AbadiABBCCCDDDG16} which is used together with Keras~\cite{chollet2015} in this paper.

\subsection{Autoencoder-Based Network Architecture}

\begin{figure}[htb]
	\centerline{\includegraphics[width=0.48\textwidth]{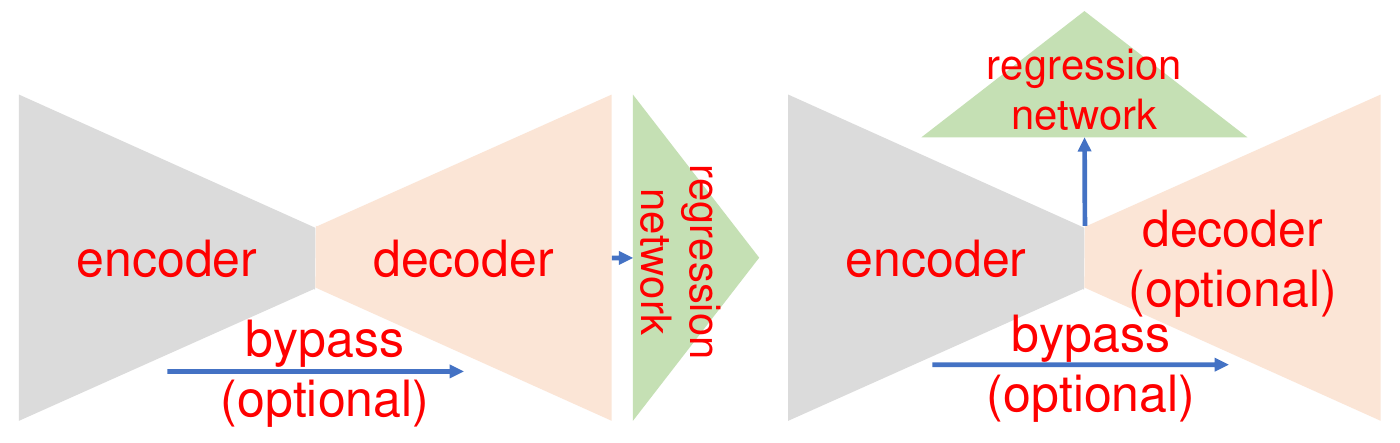}}
	\caption{Two variants of network architectures based on the autoencoder.}
	\label{fig:ae-based-nn}
\end{figure}

We propose to use the autoencoder-based network architecture~\cite{10.1007/978-3-319-24574-4_28} as our reference model, shown in \figurename~\ref{fig:ae-based-nn}. It is composed of an encoder containing convolution layers, a decoder containing deconvolution layers, an optional bypass between the encoder and the decoder, and a \emph{regression network} which can be located at the far end~\cite{Ai_2019} or at the bottleneck~\cite{Ai_2021}. The decoder is optional when the regression network is at the bottleneck. This reference model is the workload targeted by the neural network accelerator which will be discussed in Section \ref{sec:soc-acc-design}.

\subsection{Quantization-Aware Training and Validation}

\begin{figure*}[t]
	\centerline{\includegraphics[width=0.7\textwidth]{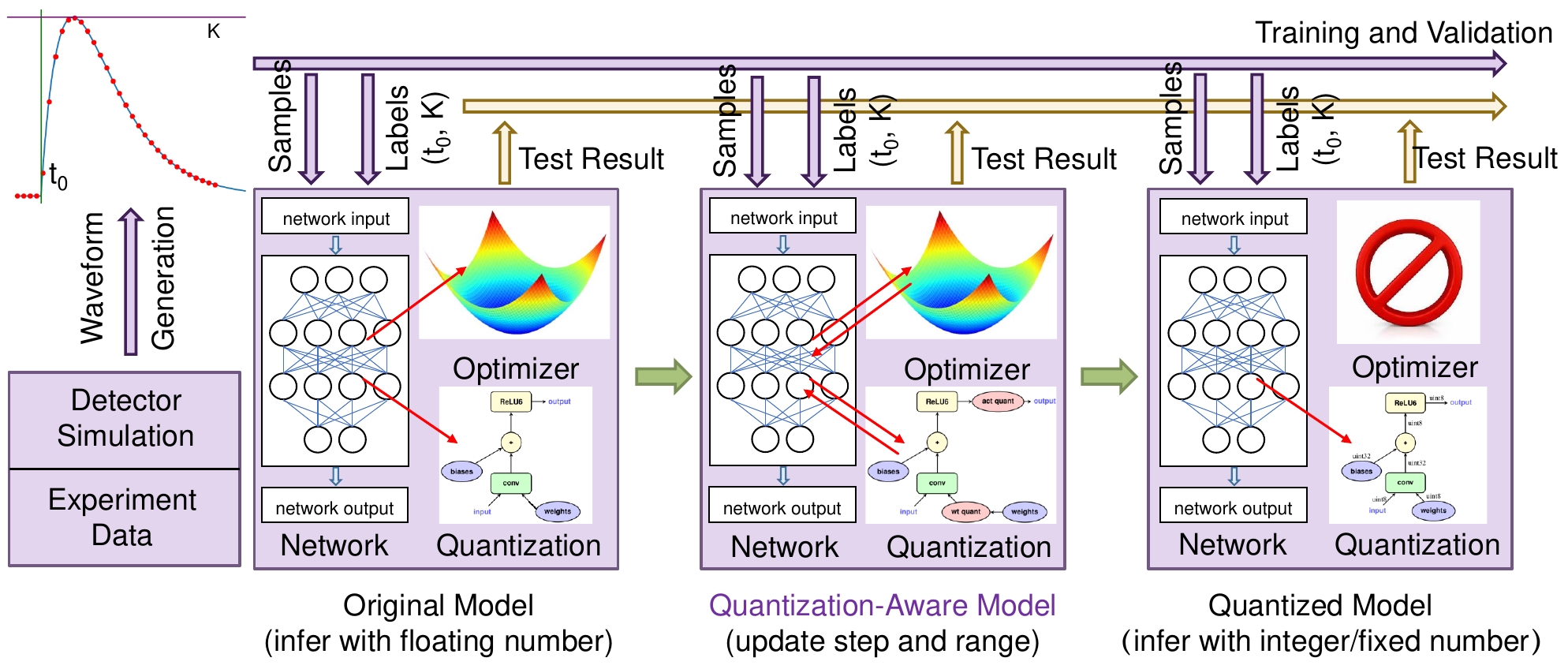}}
	\caption{Three stages of quantization-aware training. In training, labels are used to formulate the loss function. In testing, they are used to compute metrics.}
	\label{fig:quan-aware-train-val}
\end{figure*}

Conventional artificial neural networks use floating-point numbers to represent weights and intermediate feature maps, which is straightforward for CPUs and GPUs, but not friendly to customized digital ASICs. It is possible to convert the floating-point model into the fixed-point using 8-bit or 16-bit quantization. However, directly converting a well-trained model will result in significant degradation of accuracy.

To tackle the problem, a quantization-aware training scheme~\cite{8578384} natively supported by deep learning frameworks is utilized to gradually transform the original floating-point model into the quantized model with nearly no loss of accuracy. The three stages of the process is shown in \figurename~\ref{fig:quan-aware-train-val}. In each stage, waveform samples coming from detector simulation or experimental measurement are provided as network input, and signal features (such as time or energy) are provided as ground-truth labels to optimize parameters of the network and validate its performance.

The official code for quantization-aware training is assumed to be used with CPUs. To make it compatible with the neural network accelerator here, we post-process the layer-wise results with rescale and bit-shift on hardware (Section \ref{sec:acc-arch}) and rewrite the mechanism for quantized-model export in the hardware-software codesign (Section \ref{sec:hw-sw-codesign}).

\section{System-on-Chip Accelerator Design ($\beta\gamma$)}
\label{sec:soc-acc-design}

\subsection{A Brief Review of PulseDL}

For application of neural networks to process detector signals at FEE, we developed the first version of the neural network accelerator customized for pulse processing, named \emph{PulseDL}~\cite{AI2020164420,Chen2020}. \emph{PulseDL} worked as a co-processor and communicated with a RISC processor through a proprietary bus. The accelerator architecture of \emph{PulseDL} was mainly composed of a 4$\times$4 PE array, fed by dedicated row buffers and column buffers, and followed by spatial and temporal adder trees. In operation of each PE, operands from the feature map vector and the kernel matrix are multiplied and accumulated.

The first version of the chip, although a successful practice, has some notable \textbf{issues}: \textbf{(i)} a RISC CPU outside the chip (or accelerator) is needed to schedule transactions, which is not convenient in use; \textbf{(ii)} dynamic quantization (i.e., deciding the rounding bit after the feature map is obtained) may bring about additional overheads of time; \textbf{(iii)} the adder tree structure, especially the temporal adder tree, has yet to be optimized for area and performance; \textbf{(iv)} finally, only manual configuration procedure is devised to map neural network models onto the chip, while extension to deep learning frameworks could be a more automatic solution. The above limitations motivate us to develop \emph{PulseDL-II}, the new version of the digital design.

\subsection{PulseDL-II: SoC Structure}
\label{sec:soc}

\begin{figure*}[htb]
\centerline{\includegraphics[width=0.68\textwidth]{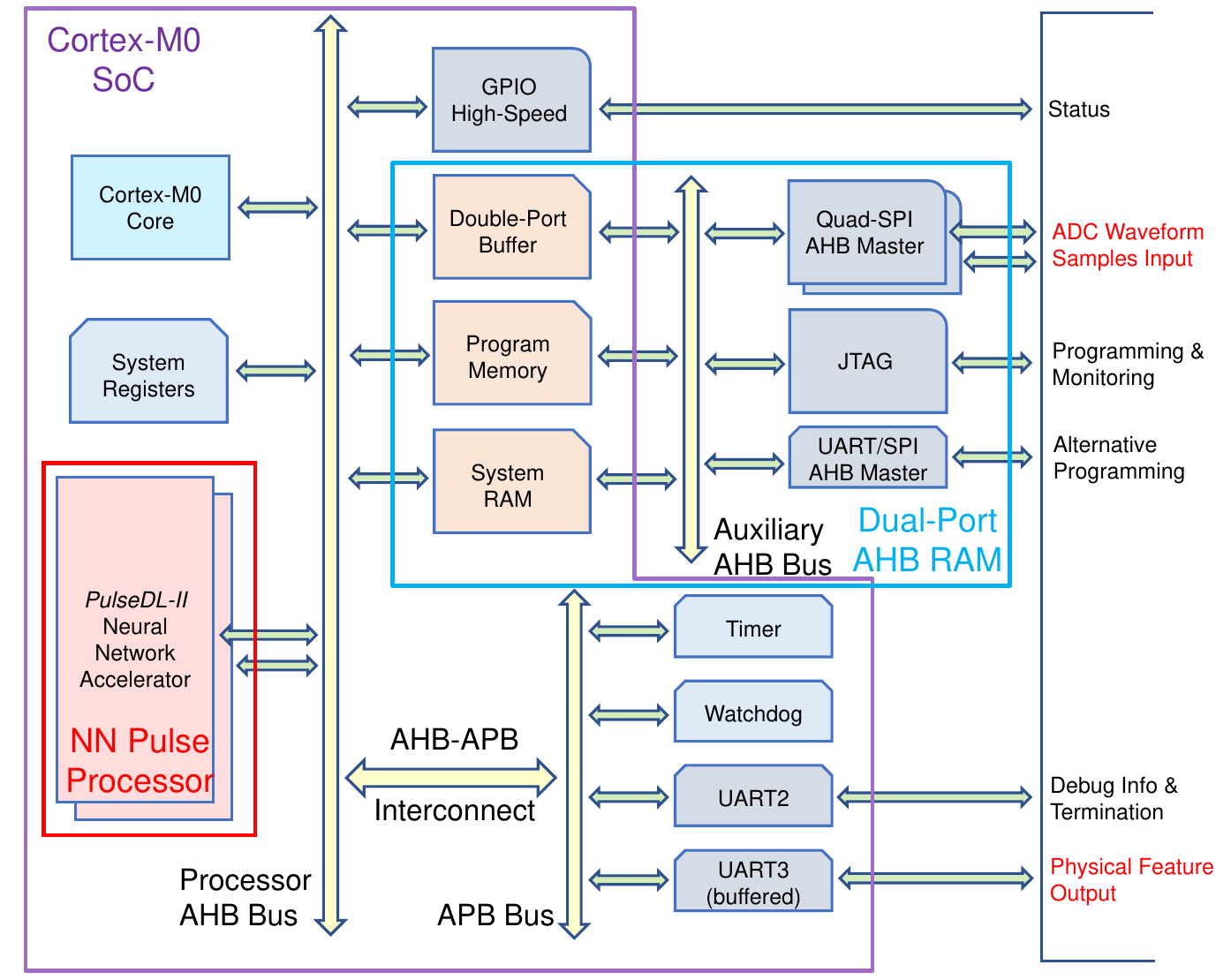}}
\caption{The system-on-chip structure of \emph{PulseDL-II}. Three major parts are surrounded by different polygons and marked with colors.}
\label{fig:soc-structure}
\end{figure*}

The primary improvement of \emph{PulseDL-II} is the system structure. We integrate an RISC CPU, the ARM Cortex-M0 microcontroller, and associated AHB/APB buses into the digital design to form an SoC, as shown in \figurename~\ref{fig:soc-structure} (which solves the \textbf{issue (i)}). The Cortex-M0 core is an intellectual property (IP) distributed freely as trial by the ARM company~\cite{arm-company}. It is a microcontroller featuring the small footprint and low power.

The SoC has three major parts: the \emph{NN Pulse Processor} (\emph{PulseDL-II} accelerator), the \emph{Cortex-M0 SoC} and the \emph{Dual-Port AHB RAM}. The \emph{PulseDL-II} neural network accelerator is mounted on the processor AHB bus as a peripheral. In this SoC, we have multiple input/output peripheral devices, such as quad/normal SPIs, UARTs (with or without the internal buffer), JTAG and GPIOs.

The preset workflow is described as follows. The waveform samples coming from ADC and self-trigger logic are transmitted into the double-port buffer, preferably through the quad SPI interface. The Cortex-M0 core periodically accesses the double-port buffer and relays the input data to the neural network accelerator. When the accelerator finishes computing, it raises a signal to the Cortex-M0 core and pushes feature maps into the system RAM under the coordination of the core. The data transmission between the system RAM and the accelerator is repeated several times, until the final feature outputs are ready. Finally, the output data are sent out through a buffered UART and collected by subsequent electronics.

It should be mentioned that transactions on the processor AHB bus generate unavoidable overheads in the total time budget (see Section \ref{sec:emb-sw-with-ws-map}). At the current stage, we have kept the CPU-centric design to maximize universality, but will probably incorporate direct memory access (DMA) with data processing abilities when we gather more information about the performance of different neural network architectures.

\subsection{PulseDL-II: Accelerator Architecture}
\label{sec:acc-arch}

\begin{figure*}[htb]
	\centerline{\includegraphics[width=0.7\textwidth]{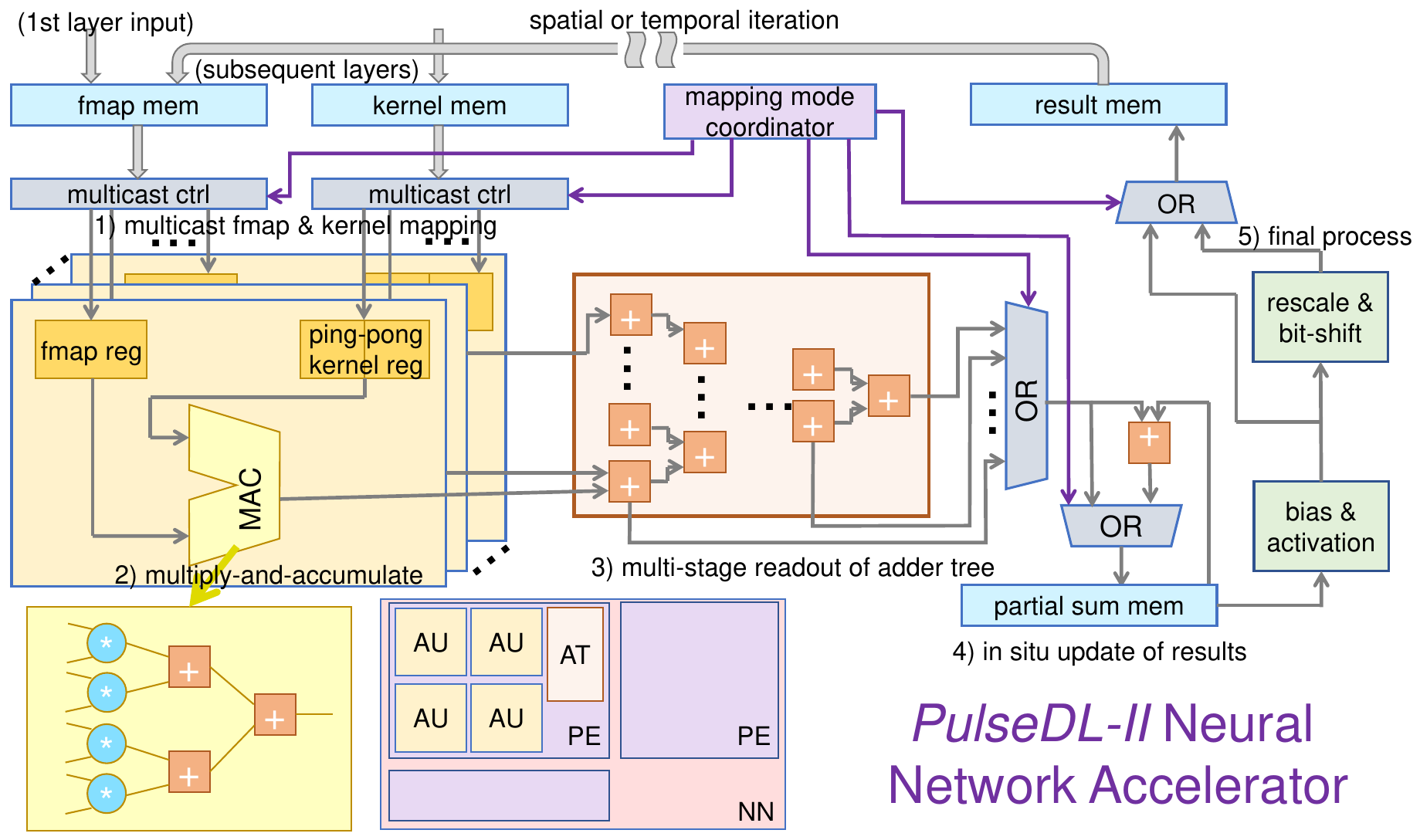}}
	\caption{The accelerator architecture of \emph{PulseDL-II}. The feature map memory interface (fmap mem) and kernel memory interface (kernel mem) receive data and multicast through controllers (multicast ctrl) to multiple arithmetic units (AU). In each AU, the feature map register (fmap reg) and ping-pong kernel register (kernel reg) store data for multiply-and-accumulate (MAC). After MAC, the adder tree (AT) sums up the outputs from AUs, and the partial sum accumulator updates results of AT in the memory (partial sum mem). Final results are transferred through the result memory interface (result mem). The square in the lower left corner shows the basic structure of MAC in an AU, and the square in the lower middle shows three hierarchical levels.}
	\label{fig:accelerator-arch}
\end{figure*}

Another important improvement is the accelerator architecture itself. The updated digital design of the \emph{PulseDL-II} accelerator is shown in \figurename~\ref{fig:accelerator-arch}. A new hierarchical level, the arithmetic unit (AU), is added into the topology. There are three hierarchical levels in total: AU, PE and NN (neural network), as shown in the middle of the bottom in this figure. In essence, PE plays a role which is more self-contained and can be regarded as a miniature \emph{PulseDL} accelerator. Accordingly, AU takes the responsibility of multiply-and-accumulate (MAC) at a more fine-grained level than the original PE in \emph{PulseDL}: instead of performing MACs with a macro PE array and post-processing the results by macro adder tree structures with lumped control logic, \emph{PulseDL-II} distributes MACs into AUs in each PE, and uses micro adder tree structures and dedicated control logic. By these functional adjustments, both efficiency and flexibility have been improved, and the software mapping scheme fits better into the hardware, which in turn reduces power and area. A \emph{template-based methodology} is adopted to design logic elements and to make elements in each level adjustable. Compared with other methods, such as high level synthesis, this methodology allows designers to perform cycle-accurate optimization and fine control of data interfaces, at the expense of more expertise and manpower. In the example design referred in the following sections, we use 4 AUs in each PE, and 15 PEs in each NN. The transactions are assumed to be directed to a single NN device.

For quantization compatible with TensorFlow or other deep learning frameworks, a functional block of rescale \& bit-shift~\cite{8578384} is integrated after temporary results are generated by bias \& activation (which solves the \textbf{issue (ii)}). Rescale \& bit-shift is a procedure to adjust the scale of the output feature map so that the full quantization bits can be effectively utilized. It serves the purpose in a way friendly to integer-only digital logic. The actual quantization bits used in the inference will determine the loss of accuracy by quantization.

In the \emph{PulseDL-II} accelerator, other enhancements include: broadcasting/multicasting input feature maps and kernels to multiple AUs, optimizing the (temporal) adder tree with the partial sum accumulator (which solves the \textbf{issue (iii)}), adding function blocks for bias addition and activation, and a whole new implementation of the mapping mode coordinator for different layers, etc.

\subsection{Hardware-Software Codesign}
\label{sec:hw-sw-codesign}

\begin{figure*}[htb]
	\centerline{\includegraphics[width=0.7\textwidth]{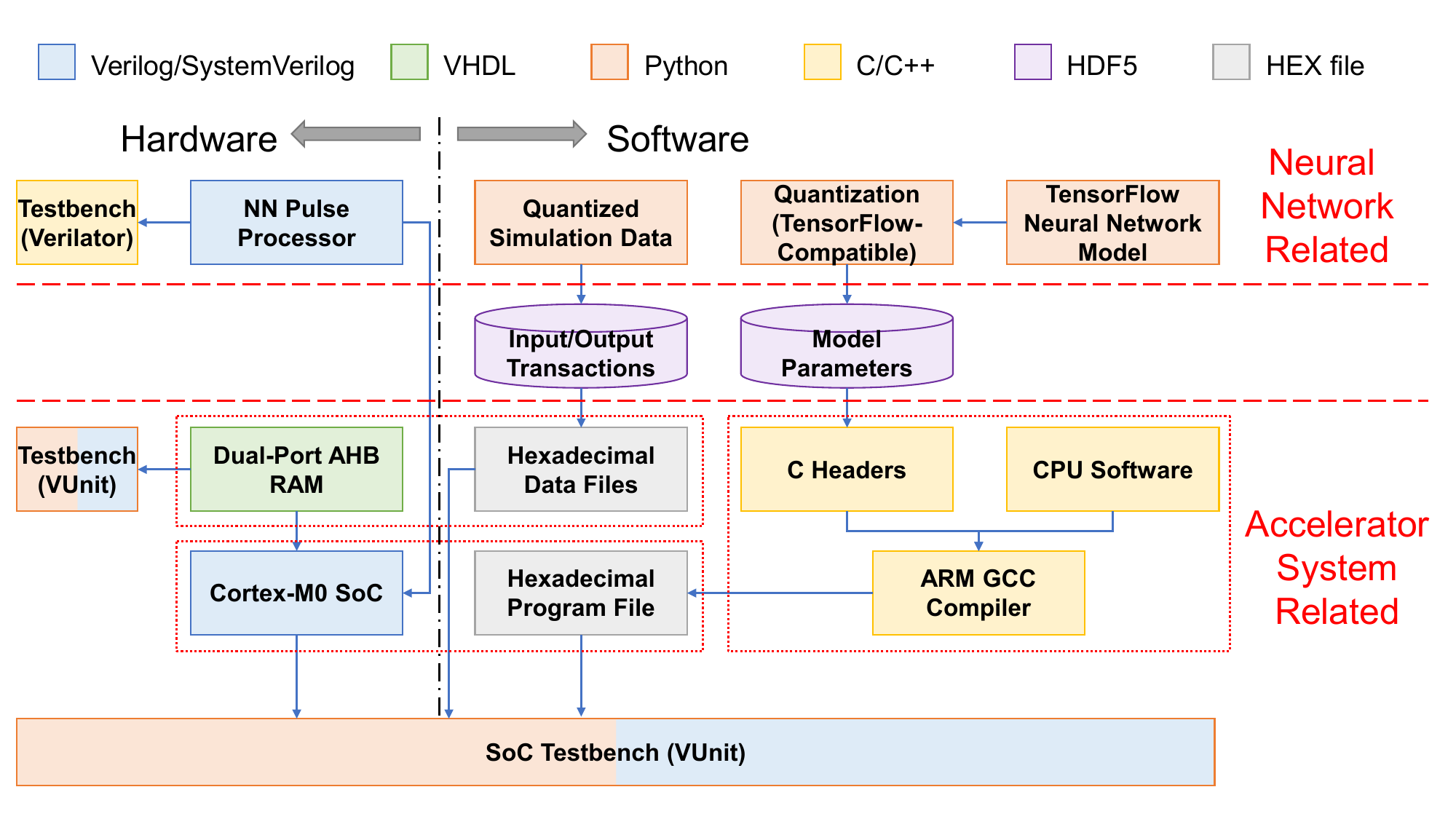}}
	\caption{The data stream diagram for hardware-software codesign of \emph{PulseDL-II}. Functional groups are divided by lines with different styles, and programming languages and file formats are indicated with colors.}
	\label{fig:hw-sw-codesign}
\end{figure*}

To ensure the proper functioning of the SoC, a framework for hardware-software codesign is implemented along with \emph{PulseDL-II}, as shown in \figurename~\ref{fig:hw-sw-codesign} (which solves the \textbf{issue (iv)}). We use the same names as \figurename~\ref{fig:soc-structure} for three major parts in the SoC, placed on the left side. The upper part in this figure is related to the neural network. Two databases, containing the input/output transactions and model parameters, are generated from neural network-related software. The lower part is related to the accelerator system. The embedded software for Cortex-M0 is compiled with the ARM GCC compiler, and converted to the hexadecimal format. At the bottom, a testbench using VUnit (a unit test framework for hardware description languages) is set up to test the function and performance of the SoC with benchmark programs and also corner cases. Multiple programming languages and file formats are used as indicated in this figure.

\subsection{Embedded Software with Weight-Stationary Mapping}
\label{sec:emb-sw-with-ws-map}

\begin{figure}[htb]
	\centerline{\includegraphics[width=0.48\textwidth]{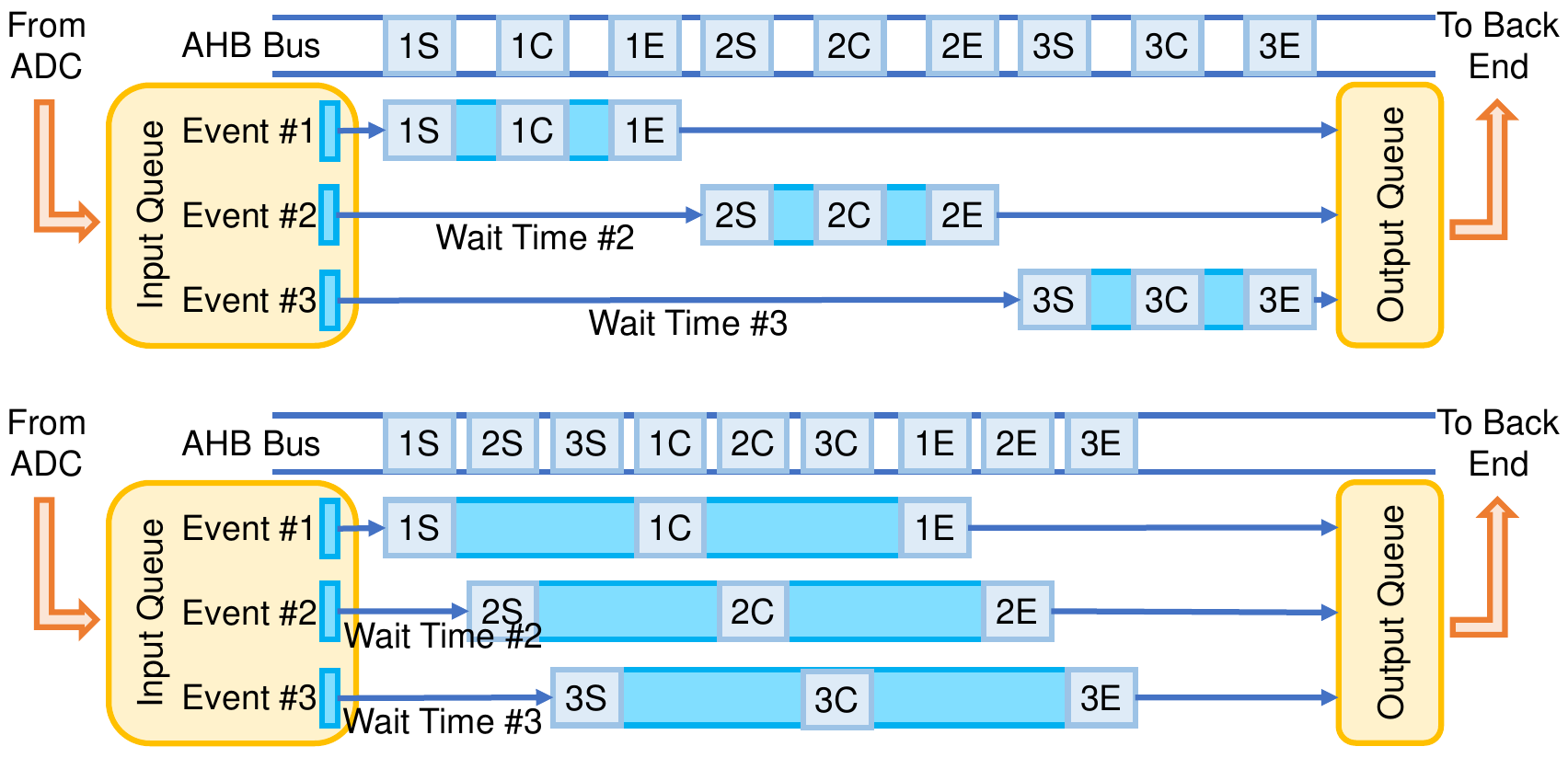}}
	\caption{The illustrated basic dataflow (top) and dataflow with pipelining and parallelism enabled (bottom).}
	\label{fig:sw-mode}
\end{figure}

Within the SoC structure, different mapping rules are allowed depending on the requirements of application. For neural networks with small/medium size, a \emph{weight-stationary} mapping scheme can be adopted to reduce transactions when weights are used repeatedly. In this scheme, the operation is divided into two phases. In the \emph{preparation phase}, weights are stored into PEs before waveform samples come in. In the \emph{inference phase}, only input data, output data and intermediate feature maps are transferred in and out of PEs. Besides, the weight-stationary embedded software enables following features:

\textbf{Layer-wise inference pipelining}: Weights for different layers are mapped to different groups of PEs; in inference, PEs can operate simultaneously as independent computing devices.

\textbf{Event-level parallelism}: Each event (In this paper, we use the terminology \emph{event} to mean a series of waveform samples generated by the detector response to a single initial particle.) is assigned a unique token and recorded by the Cortex-M0 core; in inference, the token will be traced by the core and passed in company with feature maps along the pipeline.

Fig. \ref{fig:sw-mode} shows the improvement of throughput (at the risk of prolonging latency) when above two features are enabled. The improvement is relevant to the proportion of time in data transmission and the proportion of time in neural network computation.

\subsection{Evaluation}

\begin{figure}[htb]
	\centerline{\includegraphics[width=0.2\textwidth,angle=90]{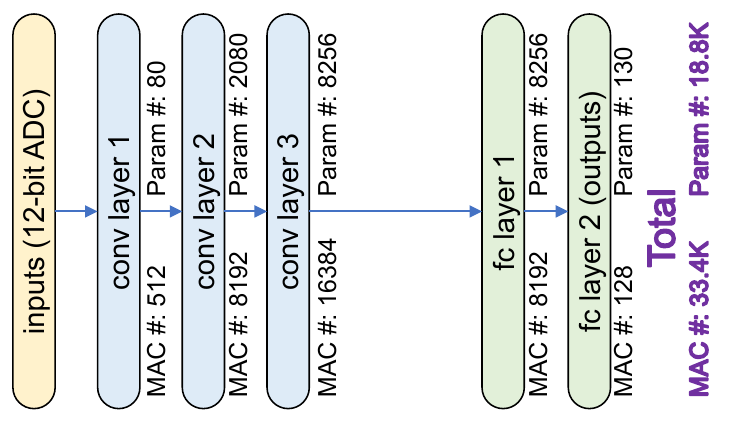}}
	\caption{Neural network workload used to compare two versions of the design.}
	\label{fig:workload}
\end{figure}

\begin{figure*}[htb]
	\subfigure[Performance (time consumption).]{
		\includegraphics[width=0.32\textwidth]{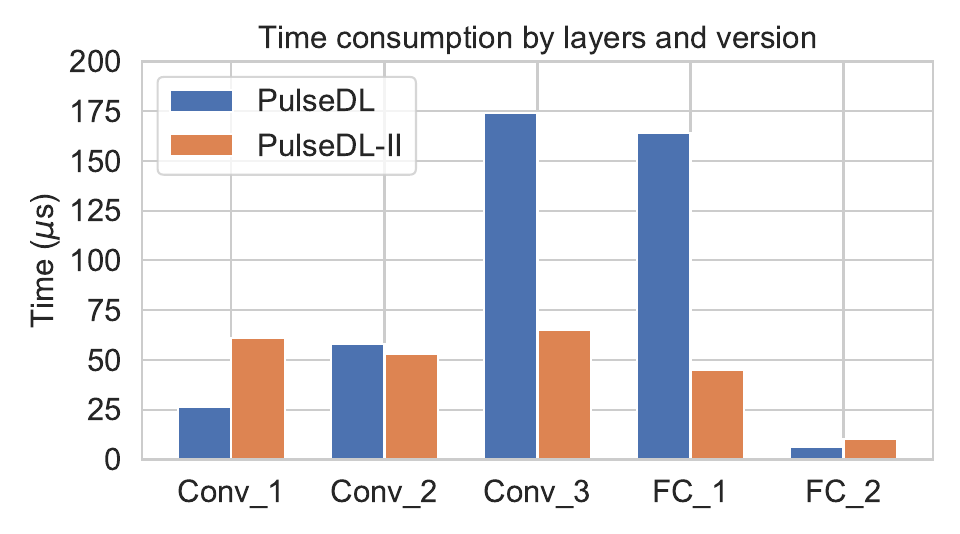}
		\label{fig:test-time}
	}
	\subfigure[Power (energy consumption).]{
		\includegraphics[width=0.32\textwidth]{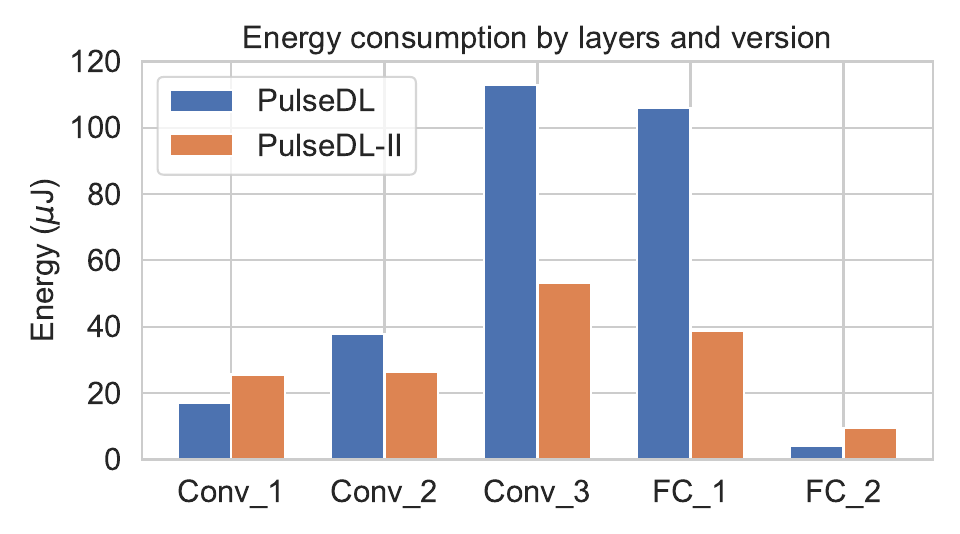}
		\label{fig:test-energy}
	}
	\subfigure[Area (resource utilization).]{
		\includegraphics[width=0.32\textwidth]{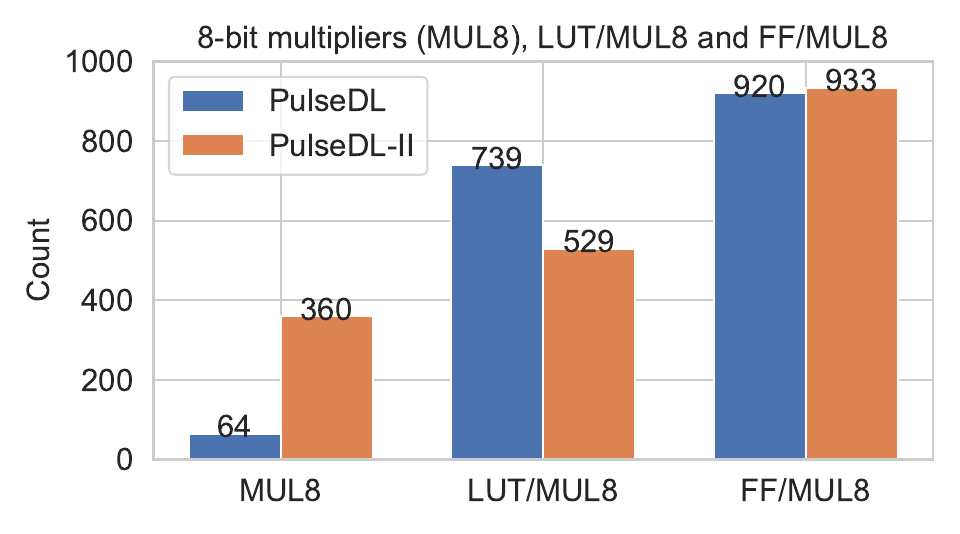}
		\label{fig:test-area}
	}
	\caption{Comparison of performance, power and area between \emph{PulseDL} and \emph{PulseDL-II} on the FPGA platform.}
	\label{fig:eval}
\end{figure*}

We compare the performance of \emph{PulseDL} and \emph{PulseDL-II} on an FPGA (Xilinx ZCU104 evaluation board) platform. The working frequency is set to 100 MHz for both versions. For fair comparison with \emph{PulseDL}, the \emph{PulseDL-II} neural network accelerator is isolated when measuring power and area.

We use a neural network workload for both versions of the hardware, comprising three convolution layers and two fully-connected layers, and containing approximately 33.4k MACs and 18.8k trainable parameters, shown in \figurename~\ref{fig:workload}.

The evaluation results are shown in \figurename~\ref{fig:eval}. In \figurename~\ref{fig:test-time} and \figurename~\ref{fig:test-energy}, it can be seen that \emph{PulseDL-II} greatly reduces running time and energy consumption. For compute-intensive layers, such as convolution layer \#3 and fully-connected layer \#1, the reduction is very significant, mainly due to the innovation in the accelerator architecture and mapping scheme. These two layers contribute largely to the overall improvement in performance (1.83$\times$ less) and power (1.81$\times$ less).

For resource utilization in \figurename~\ref{fig:test-area}, \emph{PulseDL-II} integrates more multipliers (64 versus 360), but the average utilization of FPGA hardware resources is comparable or less. The look-up tables (LUT) divided by 8-bit multipliers are decreased from 739 to 529 (1.40$\times$ less), and the flip-flops (FF) divided by 8-bit multipliers are increased from 920 to 933 (1.01$\times$ more).

The preliminary evaluation shows the advancement of \emph{PulseDL-II} for common workloads used in feature extraction of nuclear detector signals. For one thing, the new accelerator architecture, supported by the SoC structure, boosts the performance of inference; for another, power reduction is observed and resource utilization is acceptable, which makes the ASIC implementation much more promising.

\section{System Validation ($\alpha\beta\gamma$)}

\subsection{Experimental Setup}

\begin{figure}[htb]
	\centering
	\subfigure[Functional diagram.]{
		\includegraphics[width=0.48\textwidth]{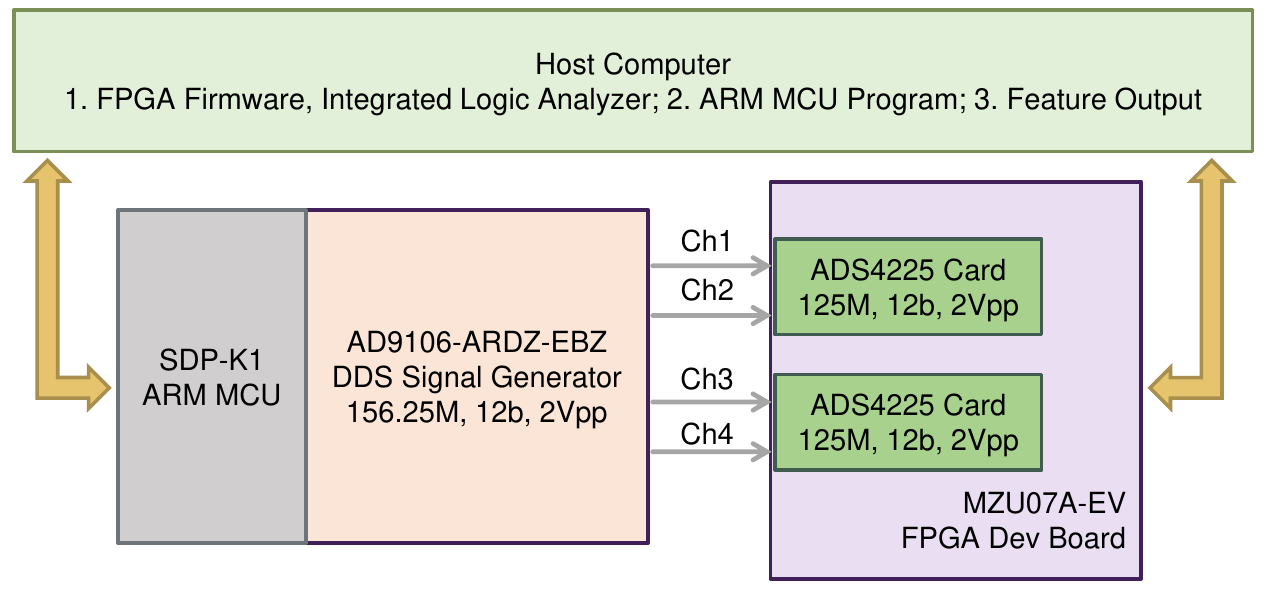}
		\label{fig:exp-setup-func}
	}
	\subfigure[Photograph.]{
		\includegraphics[width=0.3\textwidth]{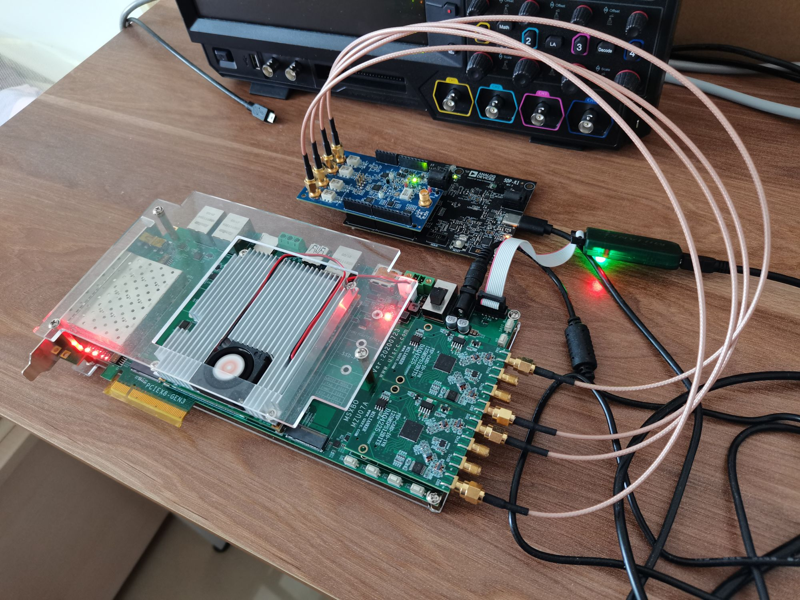}
		\label{fig:exp-setup-photo}
	}
	\caption{The experimental setup for system validation.}
	\label{fig:exp-setup}
\end{figure}

\begin{figure}[htb]
	\centerline{\includegraphics[width=0.32\textwidth]{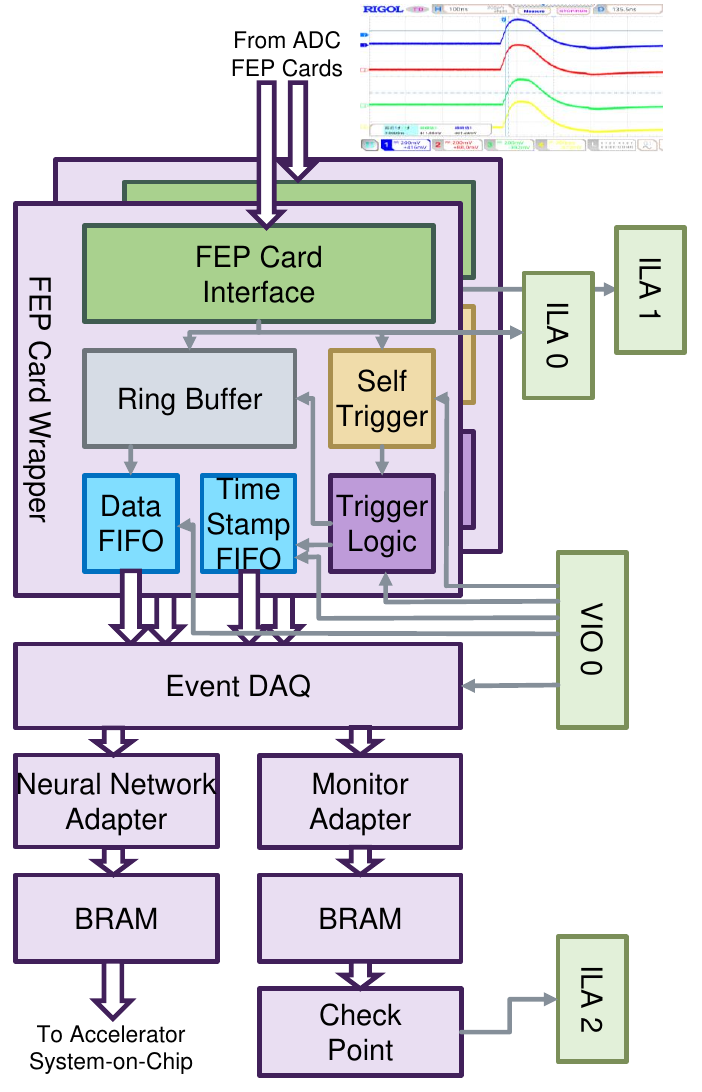}}
	\caption{The digital logic on the FPGA development board for data acquisition. The top-right square shows an example waveform from the oscilloscope.}
	\label{fig:exp-daq}
\end{figure}

To validate the whole system, we establish an experimental environment without actually connecting the electronics to the detector, as shown in \figurename~\ref{fig:exp-setup}. A DDS signal generator (AD9106-ARDZ-EBZ) working at 156.25 MHz produces 4-channel analog signals which can be programmed by an ARM microcontroller board (SDP-K1) beforehand. At the other side, we use an FPGA development board (MZU07A-EV) equipped with two ADC cards (ADS4225) sampling at 125 MHz to receive and digitize the analog signals. The data acquisition front-end and the SoC with the neural network accelerator are implemented as digital logic on the FPGA development board.

\figurename~\ref{fig:exp-daq} shows the digital logic for data acquisition, a prerequisite for feature extraction by the accelerator SoC. We use self-trigger to get a snapshot of data samples in the ring buffer and the timestamp. The triggered data is fed to event data acquisition (DAQ) for both monitoring the current waveform and preprocessing the waveform for the accelerator SoC. Multiple integrated logic analyzers (ILA) and the virtual input/output (VIO) are inserted into the dataflow to probe internal signals and manipulate the data path.

\subsection{Experimental Results}

In the experiment, we prestore the CRRC waveform (step signal filtered by the bandpass capacitance-resistance circuit) into the DDS signal generator. The waveform function is shown in Equation \ref{equ:crrc}:

\begin{equation} \label{equ:crrc}
    s(t) = K \left( \frac{t-t_0}{\tau} \right) \exp^{-(t-t_0)/\tau} u(t-t_0)
\end{equation}

\noindent where $u(t)$ is the step function. We set $\tau$ to 40 ns, and set $K = K_1 K_2$, where $K_1$ is a constant keeping the SNR relative to the baseline noise ($\sigma_{\mathrm{base}}$) to 47.4 dB:

\begin{equation}
    SNR = 20 \log_{10}\left( \frac{K_1}{\sigma_{\mathrm{base}}} \right) = 47.4\ \mathrm{dB}
\end{equation}

\noindent and $K_2$ is uniformly sampled in the range between 0.5 and 2.0. For time analysis, we produce dual-channel synchronous waveform and compare the timing results of two channels. For energy analysis, we first vary the waveform amplitude in a certain range and then use a standard waveform to assess energy predictions.

The neural network model we use is similar to \figurename~\ref{fig:workload}. It is comprised of two convolution layers and three fully-connected layers, and contains 9.8k MACs and 5.8k trainable parameters. Waveform samples from the monitoring branch in the data acquisition are used to train the neural networks and to export network weights for online processing. Identical waveform samples are used for traditional feature extraction methods.

\begin{figure}[htb]
	\centering
	\includegraphics[width=0.33\textwidth]{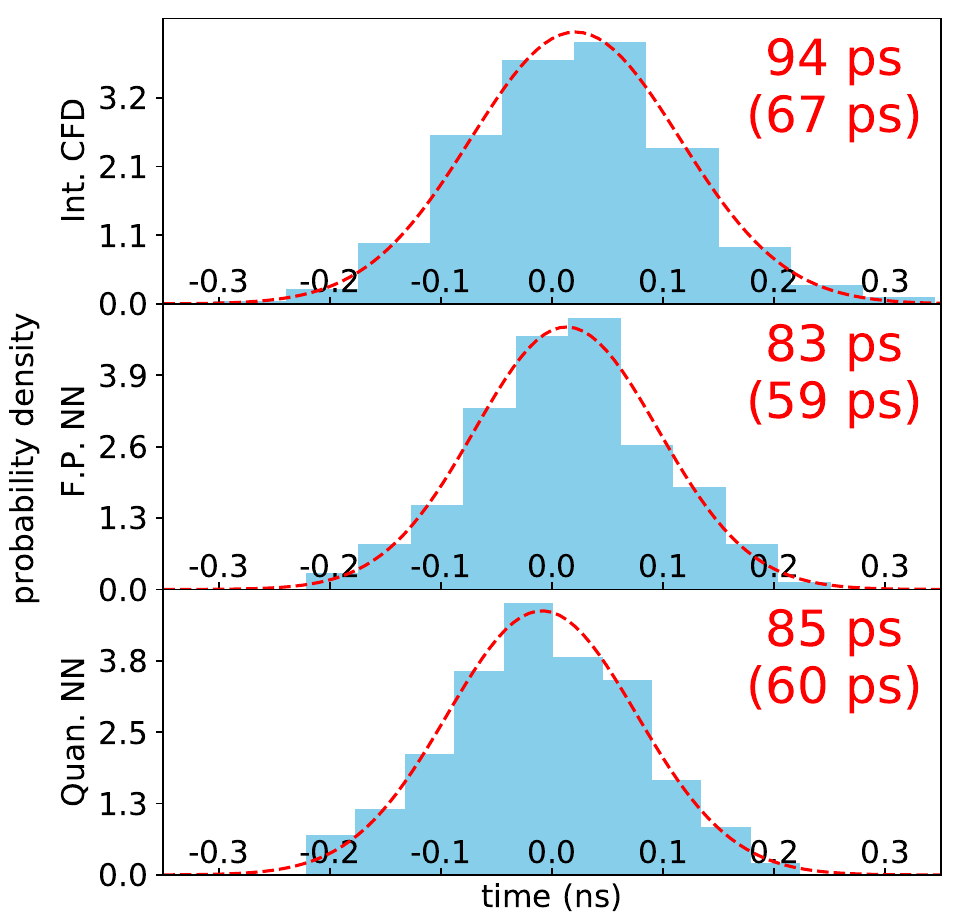}
	\includegraphics[width=0.33\textwidth]{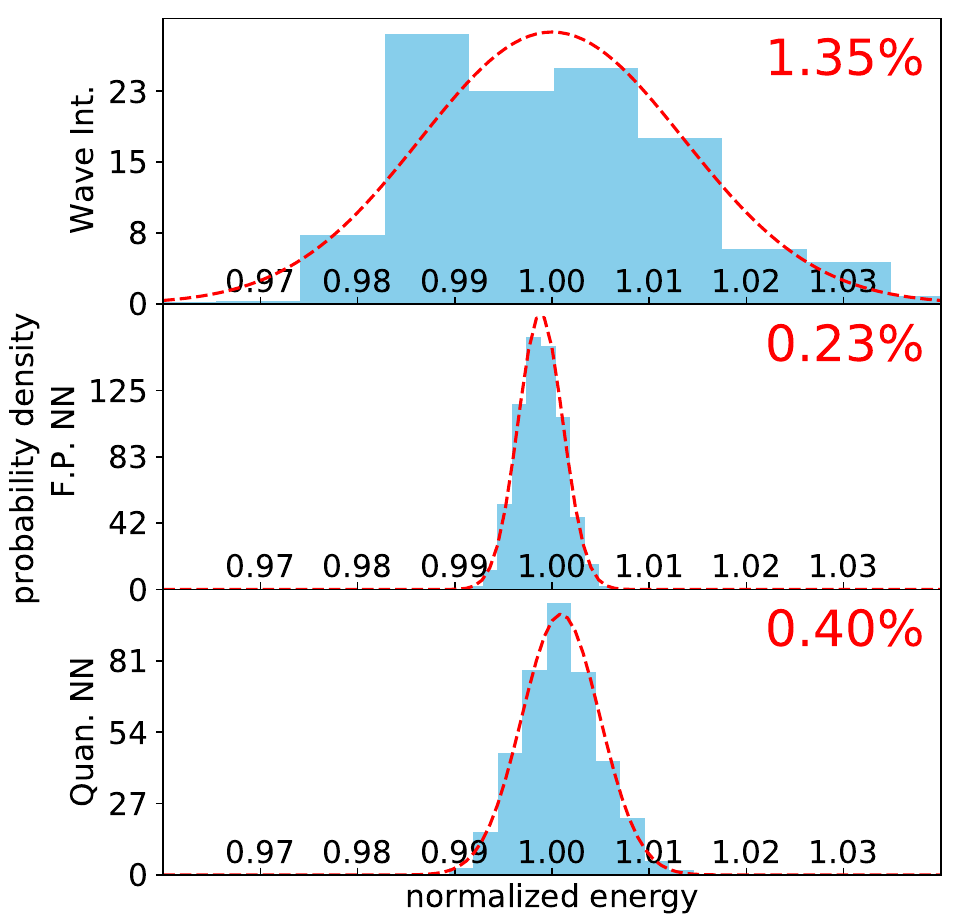}
	\caption{Experimental results of time (top) and energy (bottom) resolution by traditional methods, offline floating-point neural networks and online quantized neural networks. For time, the numbers in the parentheses are single-channel standard deviations computed from differences of two channels (divided by $\sqrt{2}$).}
	\label{fig:exp-results}
\end{figure}

In \figurename~\ref{fig:exp-results}, we show experimental results of time/energy prediction by different methods, including traditional methods (interpolated constant fraction discrimination~\cite{9187849} for time, and waveform integration for energy), offline floating-point neural networks and online quantized (8-bit fixed-point) neural networks. It can be seen that neural networks work better than traditional methods in the same conditions. Since quantization effects will influence the final online resolution, we observe a slight degradation of resolution when comparing the bottom with the middle in each sub-figure. It will be eliminated if more quantization bits are used (such as 16-bit fixed-point).

Here we report resource utilization, power and performance of the experimental system. Measured on the Xilinx Zynq UltraScale+ FPGA, the data acquisition uses 2825 LUTs, 517 FFs, 8 block RAMs (36 kb each) and 8 UltraRAMs (288 kb each), and consumes 0.371 W dynamic power. The accelerator SoC uses 89540 LUTs, 75028 FFs and 48 block RAMs, and consumes 0.541 W dynamic power. The device static power is 0.594 W. Both the dynamic power of the data acquisition and the static power are expected to improve significantly with the ASIC implementation. The time for on-chip neural network inference is 113.8 $\mu$s at 100 MHz working frequency. With time consumption from data input/output, the total latency of a single event is expected to be 165 $\mu$s. When working in the pipelined mode, the throughput is 8.3k events/second.

\section{Conclusion}

The ability and potential of neural networks in feature extraction of nuclear detector signals are investigated. Based on the advantage of neural networks, we prototype a neural network accelerator-centric FEE for ECAL at NICA-MPD. Application-specific neural network architectures are proposed, and quantization-aware training is extended for use by customized computing devices.

The major part of our work is to develop an SoC digital system with the neural network accelerator. The SoC approach is flexible not only in changing weights of a target neural network, but also in accommodating a variety of network workloads within the selected subset. We elaborate on the design of \emph{PulseDL-II} from both hardware and software aspects. A comparison between \emph{PulseDL-II} and its previous version demonstrates the advancement of digital design. Finally, system validation on an FPGA platform is done.

In the future, we will evaluate the whole system in real-world nuclear detector dataflows. We also plan to tape out with 28/65 nm process after the ASIC layout has been finished.

\bibliographystyle{IEEEtran}


\end{document}